# Isotope Effect on the Magnetic Properties of Hexamethylbenzene: Evidence of Magnetism Based on Correlated Motion of Deuterons

*Fei Yen[1*]*

[1]School of Science, Harbin Institute of Technology (Shenzhen), University Town, Shenzhen, Guangdong 518055 P. R. China

*{This manuscript is the unedited version of the Author's submitted work which was subsequently accepted (June 25th, 2018) for publication in the Journal of Physical Chemistry C after peer review. To access the final edited and published version of this work, please visit [https://pubs.acs.org/doi/10.1021/acs.jpcc.8b04255]. Thank You.}*

ABSTRACT: The associated magnetic moments of the periodic rotational motion of methyl groups in hexamethylbenzene $C_6(CH_3)_6$ were recently identified to contribute to its overall magnetic susceptibility. Those measurements however, were only performed on hydrogenous polycrystalline samples. We report magnetic susceptibility measurements on single crystalline $C_6(CH_3)_6$ in the cases where the applied magnetic field is parallel and perpendicular to the molecular basal planes. In the former case, metastable behavior near the onset temperature $T_{C\_H}$=118 K is identified while in the latter, no anomalous behavior is observed. Similar anomalies are observed in deuterated hexamethylbenzene $C_6(CD_3)_6$ (where D is deuterium), however, $T_{C\_D}$ occurs 14 K higher at 132 K. In addition, a peak anomaly identified near 42 K is suggested to be due to the onset of coherent quantum tunneling of deuterons. The near cubic ground state is proposed to be the result of a more radical form of the Jahn-Teller effect occurring in a molecular solid where the lattice distorts to remove the orbital degeneracies of the protons to lower its energy. The apparent magnetic anisotropy and isotope effect provide evidence that apart from electrons, not only protons, but also deuterons establish strongly correlated behavior.

**Introduction:**

The forms of magnetism known to date are mainly comprised of electron motion; from ferromagnetism to superconductivity and even the more exotic types such as superparamagnetism to spin glass behavior. Even the magnetism arising from molecule-based magnets is also ultimately due to unpaired electrons with extremely low onset temperatures.[1] Recently, it was suggested that the correlated rotational motion of protons ($H^+$) in the near cubic phase of the molecular solid hexamethylbenzene $C_6(CH_3)_6$ also yields a magnetic contribution[2] apart from the expected diamagnetism[3] from its electrons and 'exalted diamagnetism' from its benzoic ring currents.[4,5] If indeed periodic motions of protons comprise to exhibit magnetic behavior, new 'protonic' applications based on organic matter may be devised.

Hexamethylbenzene (HMB) is a typical molecular solid where entire $C_6(CH_3)_6$ molecules are held together by van der Waals forces. The molecule is regarded as being nearly flat with a central benzene ring with one methyl group $CH_3$ attached to each of its six vertices via C-C side bonds.[6] In the temperature range of 118–383 K, the molecules comprise a triclinic lattice structure.[7] Below 118 K, the most stable phase is a near cubic one;[8] the molecules form basal planes that extend perpendicular to the (1,1,1) direction with one exact molecule occupying its unit cell (Figs. 1a and 1b). The cohesive forces binding the molecules in HMB are mainly van der Waals,[9] from such, the periodic potential barriers inhibiting molecular motion in its solid phase are relatively low[10] that even in its bulk solid phase, entire molecules are capable of rotating and flipping.[11,12] In the case of HMB, starting from below 160 K, rotational motion of the molecules about their hexagonal axes seizes but the six methyl groups continue to rotate freely about their 3-fold axes.[11] Starting from below $T_{C\_H}$=118 K, the same temperature coinciding with the unusual triclinic to near cubic structural phase transition, various theoretical investigations[13-15] have shown how the rotation of the six methyl groups in each molecule become coupled with each other in a gear-like type of mechanism (Fig. 1c). More specifically, the three protons in each methyl group are in a pocket state, say |123> that is not in equilibrium. Oscillations to equivalent orientations, meaning simultaneously switching their positions with each other, occur at a 'tunneling' frequency, *viz.* |123> → |231> → |312> → |123> → and so on.[16] The three protons in the two adjacent methyl groups experience oscillation in the opposite direction resulting in a highly 'geared' system. Experimentally, rotational tunneling at low temperatures in HMB has been shown to occur in various studies.[11,17-22] Since the protons possess charge, their periodic rotational motion enclosing an area nearly $10^4$ times larger than electrons experiencing Larmor precession is expected to generate magnetic moments arranged in a quasi-antiferromagnetic type of spin configuration. Recently, we observed an upswing in the magnetic susceptibility at $T_{C\_H}$ which continued to increase with decreasing temperatures.[2] For the deuterated analogue of hexamethylbenzene (HMB-$d_{18}$), $C_6(CD_3)_6$, the near cubic to triclinic phase transition temperature occurs at $T_{C\_D}$=132.4 K.[23,24] To date, the only studies on the magnetic properties of HMB have been on hydrogenous polycrystalline samples. It is therefore natural to carry out similar investigations on single crystalline HMB as well as HMB-$d_{18}$ to better understand a strongly correlated system that is *not* based on electrons.

**Experimental Methods:**

Sublimed single crystalline samples of HMB >99.9% in purity were acquired from Sigma-Aldrich. The HMB-$d_{18}$ samples 99.2% in purity were acquired from CDN Isotopes. The selected samples were approximately 0.2 mm in thickness with parallelogram shapes of 3 x 3 mm and attached onto a quartz tube by GE varnish and loaded into an MPMS (Magnetic Property Measurement System) unit manufactured by Quantum Design for the magnetic susceptibility measurements. The rate at which temperature was ramped was 2–3 K/min. Error bars of the measured magnetization are omitted as their standard deviations were always over two orders of magnitude. For the case when the magnetic field $H$ was aligned perpendicular to the basal planes of the sample, two methods were employed: a) the sample was secured between two quartz stoppers held by a brass rod and measured by a PPMS (Physical Property Measurement System) unit also manufactured by Quantum Design, and b) secured onto a flat surface of a Teflon container and measured by the MPMS unit. After subtracting their background contributions, the two methods yielded similar results.

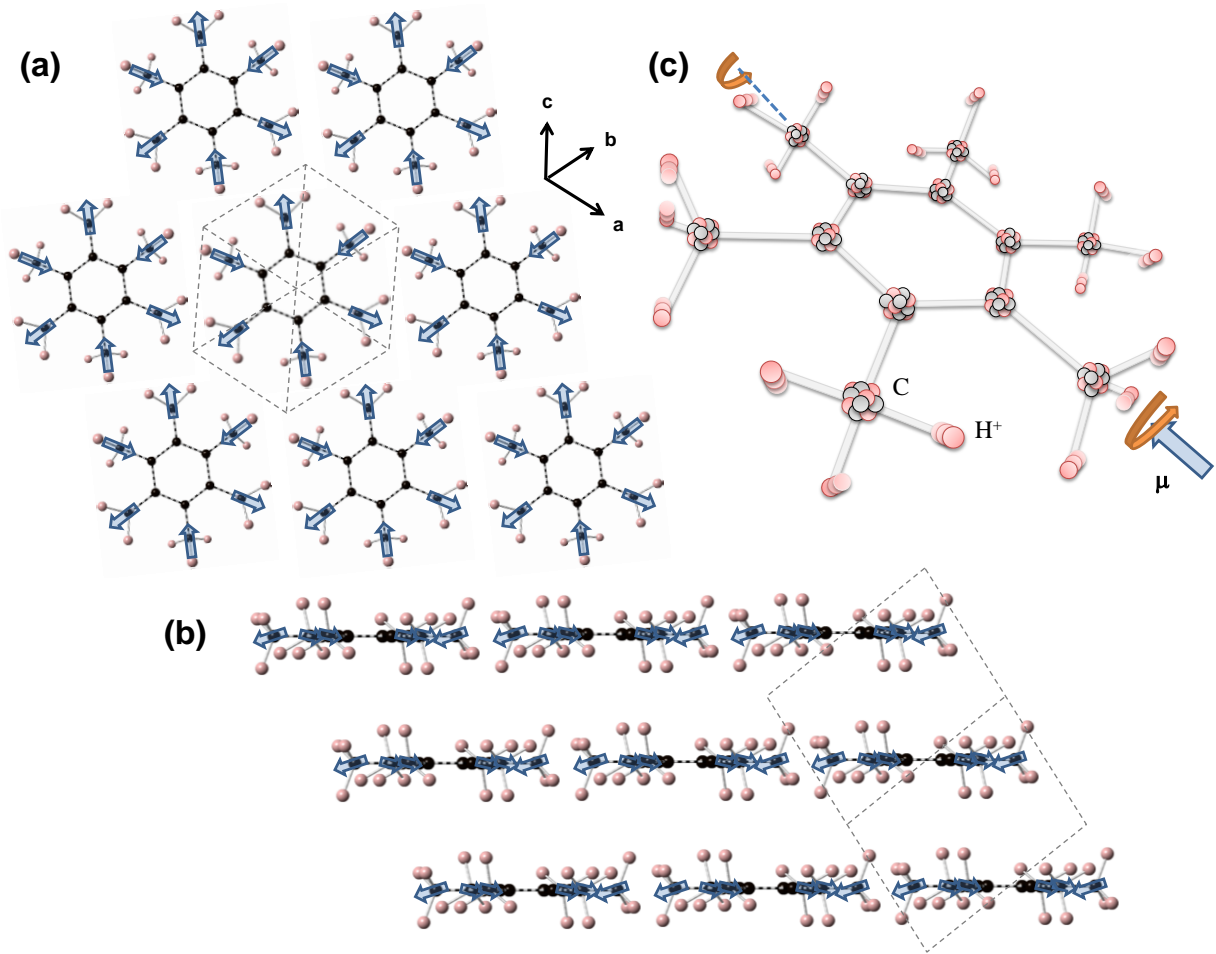

**Figure 1.** The structure of hexamethylbenzene $C_6(CH_3)_6$ (HMB). (a) The near cubic phase HMB along the (**1,1,1**) direction with basal planes forming along the perpendicular direction. (b) Cross sectional view of the basal planes. Blue arrows represent the magnetic moments **μ** associated to the gearing of the methyl groups. Grey dashed lines depict the unit cells. (c) The HMB molecule represented by protons and neutrons showing a more dynamic view of how a $\mu$ component is generated when the protons rotate within each methyl group.

**Results and Discussion:**

Figure 2a shows the magnetic susceptibility $\chi(T)$ with respect to temperature of HMB with an external magnetic field of $H$=1 kOe applied parallel to the basal plane. A step anomaly was observed in $\chi(T)$ at $T''_{C\_H}$=110.8 K during cooling and $T'_{C\_H}$=117.8 K during warming (inset of Fig. 2a). This is quite unusual since in diamagnetic systems the susceptibility is negative and independent of temperature.[3] The anomalies can be explained if a positive magnetic moment is presumed to be associated to the onset of rotational motion of the three protons in each methyl group at temperatures below $T'_{C\_H}$. In contrast to when an external magnetic field up to 10 kOe is applied perpendicular to the basal planes, the susceptibility remained nearly flat in the same temperature range (Fig. 2b) suggesting that *inter*-planar coupling is relatively small. The divergent behavior in $\chi(T)$ starting from below ~40 K for when $H$ is parallel to the basal planes (Fig. 2a and 2c) may be explained if it is taken into account that the magnetic moments of the methyl groups increase with decreasing temperature according to conclusions derived from Hartree-Foch calculations that gearing increases with cooling.[15] The weight of the magnetic moments of the methyl groups was also calculated to rival that of the electrons after taking into account their differences in mass and precession radii.[2] Furthermore, as the temperature is further lowered some of the *inter*-molecular distances between two methyl groups become smaller than adjacent *intra*-molecular distances according to inelastic neutron scattering experiments (Fig. 1a) where it was suggested that "*inter*-methyl coupling" must take place at low temperatures.[21,22] From such, the observed magnetic anisotropy herein along with the divergent behavior of $\chi(T)$ is suggestive that two-dimensional long-range ordering of the magnetic moments of the methyl groups occurs below 40 K where the associated configuration is a previously unclassified type of quasi-antiferromagnetic arrangement. Note that the magnetic moments, or 'spins', of the methyl groups do not lie on a triangular lattice such as in the case of some hexagonal rare earth manganites.[25-26] Instead, the tessellation is one that is slightly distorted and the spins are not oriented 120° from each other which appear to give rise to a high degree of magnetic frustration.

Similar to water which can exist below its melting point of 273 K in the form of metastable supercooled water or ice I$h$ all the way down to absolute zero,[27,28] the triclinic phase of HMB can also exist in metastable form below 118 K according to specific heat measurements:[29,30] when HMB was cooled past 118 K, only a hump was observed in the specific heat while upon warming a λ-shaped peak was present. Dielectric constant measurements also revealed how when HMB is cooled at 3 K/min, the system transitions from its triclinic phase to its near cubic phase at 110 K while at 2 K/min the transition occurs at 108 K, however, upon warming, the phase transition always occurred at 118 K.[2]. This explains the step anomalies in $\chi(T)$ during warming and cooling of being inherent of the system with the variations in the critical temperatures upon cooling arising from metastable behavior.

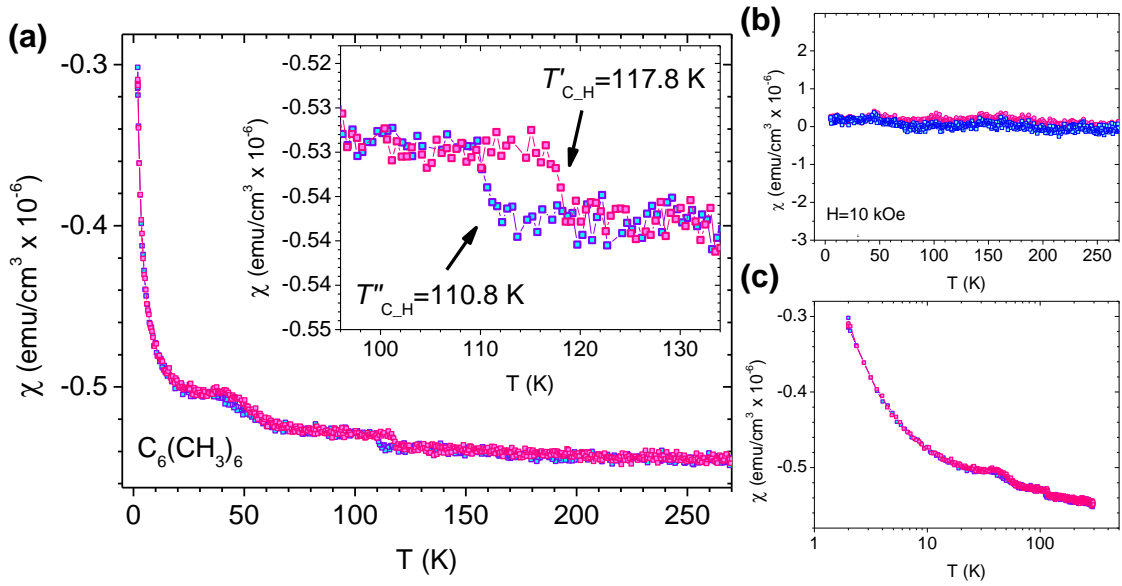

**Figure 2.** (a) Volume susceptibility $\chi(T)$ of crystalline hexamethylbenzene $C_6(CH_3)_6$ with respect to temperature under an external magnetic field of $H$=1 kOe oriented along the direction parallel to the basal planes during cooling and warming. Inset is an enlargement of $\chi(T)$ at the near cubic to triclinic phase transition. (b) $\chi(T)$ with $H$ applied along the direction perpendicular to the basal planes. (c) $\chi(T)$ portrayed in logarithmic scale.

Figure 3 shows $\chi(T)$ of HMB-$d_{18}$ at $H$=5 kOe. Curves at lower $H$ values exhibit the same features but with more noise since the samples were smaller; an $H$=2 kOe curve is also provided in the supporting information section. Similar to HMB, step anomalies were observed at $T''_{C\_D}$=126.7 K and $T'_{C\_D}$=132.2 K in HMB-$d_{18}$ which are in excellent agreement with the reported isotope effect observed in Raman and near-infrared spectroscopy[23] and specific heat measurements.[24] Since the electronic structures of both isotopes are identical, the observed isotope effect essentially pinpoints the magnetic anomalies in HMB (and HMB-$d_{18}$) to stem from the onset of correlated proton (and deuteron) motion. In addition, a peak anomaly is observed near $T_{p\_D}$=42 K in HMB-$d_{18}$. For the case of HMB, $T_{p\_H}$=40 K but the feature is less pronounced. According to NMR measurements the tunneling frequency of HMB below 20 K is approximately 10 MHz,[18] which accordingly, the system should condense into its ground tunneling state near below 40 K. Our results suggest that the most likely scenario taking place is that above $T_p$ the system possesses excited tunneling states which allow the deuterons to go over the potential barriers, whereas below $T_p$ the deuterons undergo coherent quantum tunneling straight through the potential barriers.

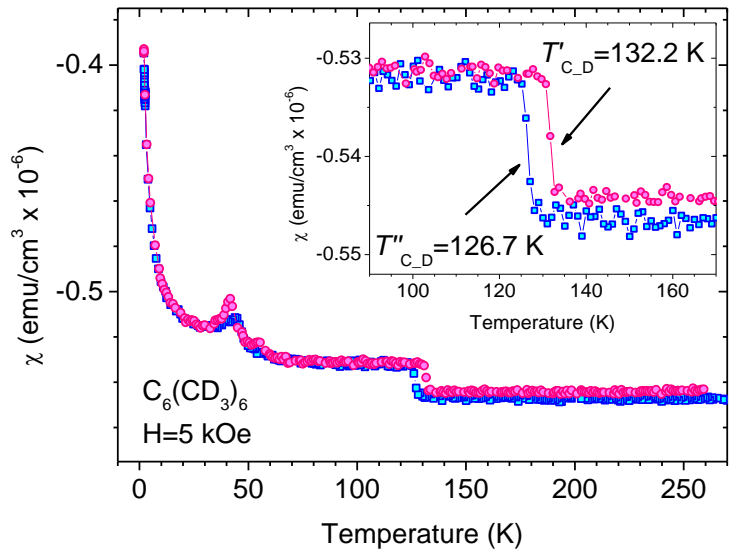

**Figure 3.** $\chi(T)$ of crystalline deuterated hexamethylbenzene $C_6(CD_3)_6$ with $H$=5 kOe oriented parallel to the basal planes.

The deuterated analogues of hydrogen based molecular solids usually exhibit a slightly higher critical temperature as the former possesses a more structured lattice. Taking for instance ice ($H_2O$) and heavy ice ($D_2O$), the difference in their critical temperatures is usually 3-4 K apart.[31,32] For the present case, the difference in the transition temperature of 14 K between the two isotopes is anomalously large. Given that the carbon skeletons of the two isotopes are nearly identical, it is apparent that the onset of magnetic ordering is extremely sensitive to the degree of compactness of the methyl groups. It appears that the transition from the triclinic to the near cubic phase (technically still triclinic) is likely a consequence of magnetic ordering of the methyl groups at the molecular level inducing a structural phase transition rather than associated magnetic moments accompanying a supposed structural phase transition due to a lower Gibbs free energy. The exact underlying mechanism remains to be investigated, however, one likelihood is the following: For protons to tunnel in concert their wave functions must be the same; meaning that they must possess the same quantum state. When the methyl groups become coupled at 118 K, to remove the orbital degeneracies of the protons, the lattice distorts itself in order to lower its energy, a sort of Jahn-Teller effect. In this case, the molecular lattice is mostly held by van der Waals forces which are typically an order of magnitude lower than Coulombic forces resulting in a more radical form of distortion. The unique set of lattice constants of the near cubic phase ($a$=6.1803, $b$=6.190, $c$=6.1993 for $C_6(CD_3)_6$ at 5 K)[8] appears to be a delicate balance between magnetic frustration and van der Waals forces. More thorough studies on what appears to be a previously unclassified type of spin-lattice coupling and Jahn-Teller distortion based on methyl group spins and van der Waals forces is obviously needed. Nevertheless, application of hydrostatic pressure to HMB, which will reduce *inter-* and *intra*-molecular distances, should enhance its magnetic properties and possibly bring forth interesting phenomena due to the hardening of its geometric frustration.

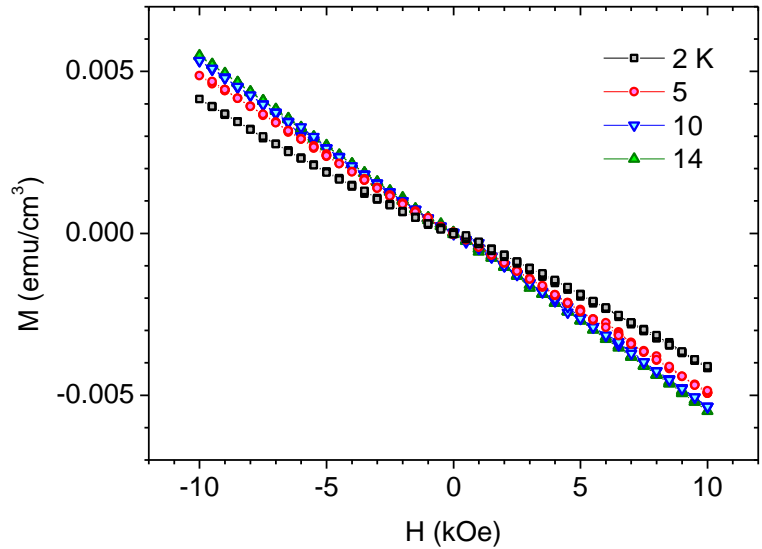

**Figure 4.** Magnetic field dependence of the volume magnetization $M(H)$ of HMB with $H$ parallel to the basal planes.

Figure 4 shows the dependence of the volume magnetization $M(H)$ with respect to $H$ oriented along the basal planes at 2, 5 10 and 14 K for HMB up to ±10 kOe. The $M(H)$ and $\chi(T)$ are in excellent agreement with each other. For a purely diamagnetic system, the $M(H)$ curves should be linear and identical at all temperatures. The nonlinear behavior of the 2 K curve is more evidence that magnetic coupling exists between the magnetic moments of neighboring methyl groups. Extrapolating $\chi(T)$ in Fig. 2c to low temperatures, it is found that below 800 mK, the susceptibility becomes positive suggesting that the magnetic contribution from the orbital motion of the protons within their methyl groups is expected to surpass the diamagnetism originating from the electrons. From such, the slope of the $M(H)$ curves is expected to become positive below ~800 mK.

**Conclusion:**

In conclusion, the structural phase transition between the triclinic and near cubic phases of hexamethylbenzene (HMB) at 118 K and deuterated hexamethylbenzene (HMB-$d_{18}$) at 132 K is believed to stem from protons and deuterons ordering magnetically at the molecular level, respectively. The unique set of lattice constants of the near cubic phase appears to be a result of the system lowering its energy in order to remove the orbital degeneracies of the protons (or deuterons in the case of HMB-$d_{18}$) via Jahn-Teller distortions. Coherent quantum tunneling of protons appears to occur below 40 K in HMB and 42 K in HMB-$d_{18}$. The existence of magnetic anisotropy is in agreement with the fact that the associated magnetic moments lie mainly along the basal planes and that *inter*-planar coupling is limited. For the case of HMB-$d_{18}$, concerted motion of 36 particles (18 protons and 18 neutrons) requires their wave functions to be the same suggesting the formation of composite bosons at quite high a temperature making it highly attractive for the study of many-body tunneling. The rotational motion of the protons and deuterons is equivalent to a rotor assembly which may be feasible to constructing molecular machinery controllable by applied magnetic fields. Lastly, many other poly-methylbenzenes also experience rotational tunneling at low temperatures.[11,33] An even larger number of compounds containing $NH_4^+$, $NH_3$ and $CH_4$ (which possess larger rotational constants) also exhibit rotational tunneling.[16] The subfield of "strongly correlated proton systems" is one that remains largely unexplored.

ASSOCIATED CONTENT

**Supporting Information**.

The following files are available free of charge.

Figure S1: Volume susceptibility of $C_6(CD_3)_6$ with $H$=2 kOe (PDF).

Figure S2: The warming curves of Figs. 2a and 3 plotted together for comparison (PDF).

AUTHOR INFORMATION

**Corresponding Author**

*Tel: +86-133-4929-0010. Email: fyen@hit.edu.cn

**Funding Sources**

The National Natural Science Foundation of China grant numbers 11374307 and 11650110430.

**Notes**

The author declares no competing financial interests.

ABBREVIATIONS

HMB, hexamethylbenzene $C_6(CH_3)_6$; HMB-$d_{18}$, deuterated hexamethylbenzene $C_6(CD_3)_6$; NMR, Nuclear Magnetic Resonance.

ACKNOWLEDGEMENTS

This work was supported in part by the National Natural Science Foundation of China grant numbers 11374307 & 11650110430.

TOC Graphic

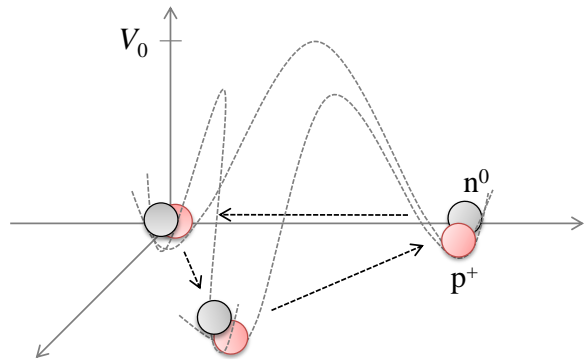

SUPPORTING INFORMATION

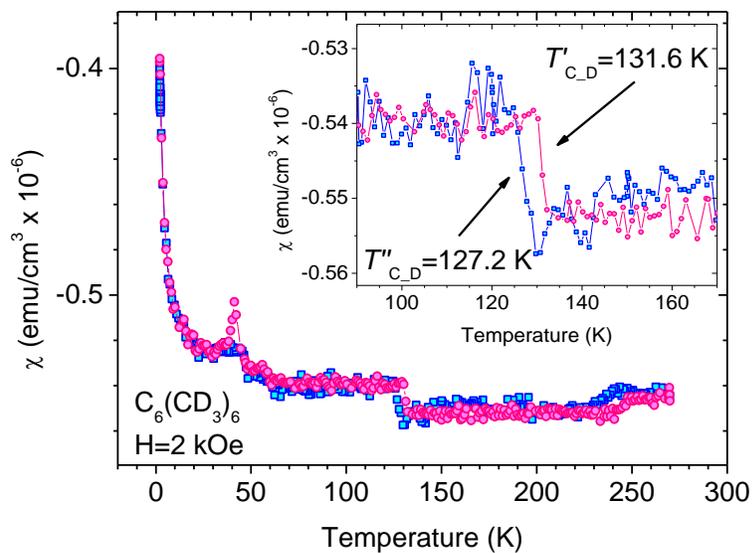

**Figure S1.** Volume susceptibility of crystalline deuterated hexamethylbenzene $C_6(CD_3)_6$ where D is deuterium with *H*=2 kOe oriented parallel to the basal planes.

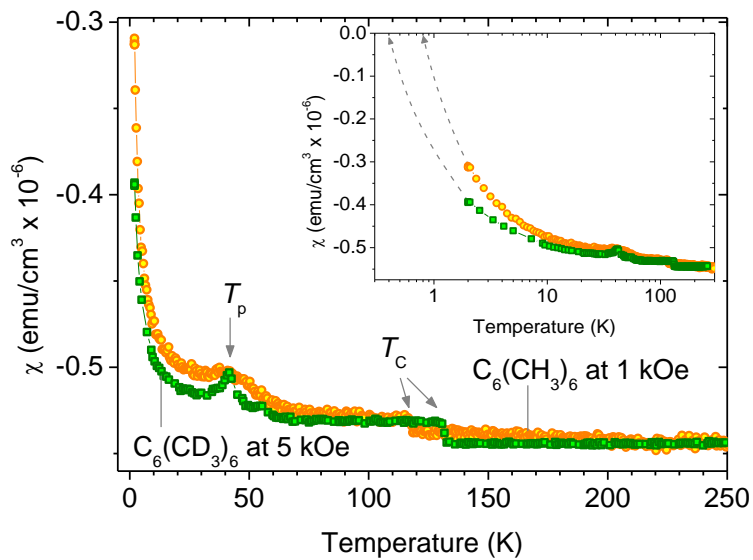

**Figure S2.** The warming curves from Fig. 2a and 3 plotted together for comparison. Inset is the low temperature region plotted in logarithmic scale. Dashed arrows are extrapolations to when $\chi(T)=0$. For $C_6(CH_3)_6$, $T \approx 800$ mK and $C_6(CD_3)_6$, $T \approx 400$ mK.